\documentclass[aps,prb,twocolumn,showpacs]{revtex4}
\usepackage{graphicx}

\begin{document} 
\title{ Spin-precession vortex  and spin-precession supercurrent stability \\ in $^3$He-B}  
\author{ E.B. Sonin}

\affiliation{ Racah Institute of Physics, Hebrew University of
Jerusalem, Jerusalem 91904, Israel} 

\date{\today} 

\begin{abstract}
The Letter analyzes stability of spin precession currents in superfluid $^3$He-B  when the precession angle is very close to 104$^\circ$. In this limit a  spin-precession vortex has a very large core, and  a barrier blocking motion of  these large-core vortices across the current streamlines (phase slip) disappears at precession-phase gradients much smaller than critical gradients  estimated from the Landau criterion. Nevertheless, spin precession currents remain stable up to the Landau critical gradients, since in this case there is a barrier, which blocks the phase slip at very early stage of vortex-core nucleation. The Letter also predicts a second-order phase transition between a parity-symmetric and parity-nonsymmetric spin-precession vortex cores at the precession angle 126.5$^\circ$. 
\end{abstract} 

\pacs{67.30.hj, 67.30.he}

\maketitle

The phenomenon of spin superfluidity was intensively studied experimentally and theoretically in 70s and  80s of the last century (see the review articles \cite{Usp,Fom-91,Bun} and references therein).
Nowadays there is a revival of interest to the phenomenon of spin superfluidity in $^3$He-B \cite{n,n1,n2,n3,n4}.  Efforts to observe a similar phenomenon   in $^3$He-A were also reported\cite{A}. Meanwhile, there still remains some unresolved controversy, which was inherited from the previous rounds of studying this problem: what is the structure of spin-precession vortex \cite{fn}, which participates in the phase-slip process, and what determines  stability of the spin-precession supercurrent in $^3$He-B.

The concept of  spin vortex as a phase-slip tool determining stability of the superfluid spin current was introduced in 1978\cite{son-78}. For the superfluid  $^3$He-B the spin-precession vortex was considered in 1987\cite{son-87} (see also Ref.~\cite{son-88}). It was obtained  that the core radius $r_c$ of the vortex is on the order of the dipole length:  $r_c \sim \xi_d =c_\perp /\Omega$, where $\Omega$ is the longitudinal-NMR frequency and $c_\perp$ is the velocity of transversal spin waves. Using the Landau criterion it was shown that the critical phase gradient is also determined by the inverse dipole length. The barrier  for vortex growth in the phase-slip process vanishes at phase gradients of the order of the inverse core radius. So the threshold for vortex instability agrees with the critical gradient in the Landau criterion. This is usual in the superfluidity theory \cite{Usp}. 

One year later Fomin\cite{Fom-88} suggested that the vortex core must be determined by another scale  $\xi_F=c_\perp/\sqrt{(\omega_P-\omega_L)\omega_L}$, where  $\omega_P$ and $\omega_L$ are 
 the precession and the Larmor frequencies.  This was supported by Misirpashaev and Volovik\cite{misir} on the basis of the topological analysis. 
Since  $1/\xi_F^2$ plays a role of the chemical potential for the precession moment and is directly connected with $\xi_d$, the question whether the core radius is determined by $\xi_d$ or $\xi_F$, is similar to the question whether the core radius in the bose-liquid is determined  by the liquid density, or by the chemical potential:  Both answers are correct since the quantities are connected by thermodynamic relations.  A real important difference was  that according to Fomin if the precession angle $\beta$ approaches to the critical value $\beta_c=1.82$ rad (or $104^\circ$)  the core radius becomes  $r_c \sim \xi_F \sim \xi_d/(\beta-\beta_c)$, i.e., by the large factor $1/(\beta- \beta_c)$ differs from the estimation $r_c \sim \xi_d$ done in Ref. \cite{son-87}. So the latter is valid only far from the critical angle, where $\beta- \beta_c \sim 1$. 

Since no barrier impedes vortex expansion across a channel if the gradient is on the order of $1/r_c$, the large core $r_c \sim \xi_d/(\beta-\beta_c)$ at  $\beta \to  \beta_c$   leads to the strange (from the point of view of the conventional superfluidity theory) conclusion: The instability with respect to vortex expansion occurs at the phase gradients $\sim 1/r_c$ essentially less than the Landau critical gradient $\sim 1/\xi_d$, obtained in Ref. \cite{son-88} for {\em any} $\beta>\beta_c$. The present Letter suggests resolution of this paradox. It demonstrates that at precession angles close to 104$^\circ$ at phase gradients less than the Landau critical gradient but larger than the inverse core radius  
no barrier impedes phase slips at the stage of vortex motion across streamlines, but there is a barrier, which blocks phase slips on the very early stage of nucleation of the vortex core. So for these gradients stability of current states is determined not by vortices but by vortex-core nuclei.

The analysis addresses also possible symmetries of the vortex core. It was expected that parity symmetry (its definition is given below) is always broken\cite{misir}. The present Letter presents numerical calculation demonstrating the second-order transition between a parity-symmetric and a parity-nonsymmetric vortex at the precession angle 126.5$^\circ$.  In the past  the first-order transition in  cores of $^3$He-B mass vortices  was  detected in NMR experiments on rotating $^3$He-B \cite{PhTr}.
It was theoretically explained in Refs.~\cite{Thun,Vol} in terms of the transition between the axisymmetric and non-axisymmetric cores. Later this theory was confirmed experimentally by direct observation of the non-axisymmetric core in one of the two vortices \cite{Core}.

The spin dynamics of superfluid phases of $^3$He is described by the theory of Leggett and Takagi\cite{LT}. Following Fomin \cite{Fom-91} we introduce the Euler angles $\alpha$, $\beta$, and $\gamma$ in the spin space of the $^3$He-B order parameter. The angle $\beta$ is the precession angle, and $\alpha$ is the precession phase. The angle $\Phi=\alpha+\gamma$ characterizes the resultant rotation of the order parameter in the laboratory frame, and in the limit $\beta\to 0$ (no precession) becomes the angle of rotation about the $z$ axis. The moments canonically conjugate to the angles $\alpha$, $\beta$, and $\Phi$ are respectively: $P=M_z-M_\xi$, $M_\beta$, and $M_\xi$, where $M_z$ is the $z$ component of the magnetization $\vec M$ in the laboratory coordinate frame, $M_\xi$ is the projection of $\vec M$  on the $\xi$ axis of the rotating coordinate frame, and $M_\beta$ is the projection of $\vec M$  on the axis perpendicular to the axes $z$ and $\xi$. 

For phenomena observed experimentally only  one degree of freedom is essential, which is connected with the conjugate pair ``precession phase $\alpha$--precession moment $P$''.  The Hamilton equations for the precession mode are: 
\begin{eqnarray}
{\partial \alpha\over \partial t}=\gamma {\delta F\over \delta P},
          ~~{\partial P\over \partial t}=-\gamma {\delta F \over \delta \alpha } . 
    \label{prec}      \end{eqnarray}   
Since the degree of freedom connected with the conjugate pair $M$--$\Phi$ is not active, the angle $\Phi $ is determined from minimization of the energy: $\delta F /\delta \Phi=0$. 
The free energy $F=F_Z+F_\nabla +V$ includes the Zeeman energy $F_Z=-\vec M\cdot \vec H=-M H u=-\chi \omega_L^2/\gamma^2$, where $\vec H=H\hat z$ is an external constant magnetic field,
the gradient energy (we assume that the spin current is normal to the magnetic field  $H\hat z$), 
\begin{eqnarray}
F_\nabla={\chi c_\perp^2 \over \gamma^2}\left[A(u){\nabla\alpha^2\over 2}
+{c_\parallel^2\over c_\perp^2}{\nabla\Phi^2\over 2}+{\nabla u^2\over 2(1-u^2)}
\right],
          \end{eqnarray}   
where
\begin{eqnarray}
A(u)={c_\parallel^2\over c_\perp^2} (1-u) ^2+1-u^2,
          \end{eqnarray}   
and the dipole energy $V ={\chi c_\perp^2 v(u,\Phi)/ \gamma^2 \xi_d ^2}$, where
\begin{eqnarray}
v(u,\Phi) ={2 \over 15 }\left[(1+\cos \Phi )u +\cos \Phi -{1\over 2}\right]^2.
    \label{DE}           \end{eqnarray}   
Here  $\chi$ is the magnetic susceptibility, $\gamma$ is the gyromagnetic ratio,  $\omega_L=\gamma H$ is the Larmor frequency,  $u=\cos \beta$,   and the  ratio $c_\parallel/c_\perp$ of  velocities of longitudinal and transversal spin waves will be chosen to be $\sqrt{4/3}$\cite{Fom-88}. In the state of stationary precession the precession angular velocity is constant: $\partial \alpha/\partial t=-\omega_P$. This state corresponds to the extremum of the Gibbs thermodynamic potential, which is obtained from the free energy with the Legendre transformation $G= F+\omega_P P/\gamma$. Thus the precession frequency $\omega_P$ plays the role of the ``chemical potential'' conjugate to the precession moment density $P$.  The distribution of the parameters $u=\cos\beta$ and $\Phi$ is determined from the two Euler-Lagrange equations $\delta G /\delta u=0$ and  $\delta G / \delta \Phi=0$.

For uniform precession minimization with respect to $u$ yields the relation \begin{eqnarray}
{\chi(\omega_P-\omega_L)\omega_L\over \gamma^2} +{\partial V\over \partial u}
\propto {1\over \xi_F^2}+{1\over \xi_d^2}{\partial v\over \partial u}=0,
       \label{u-phi}   \end{eqnarray}   
and minimization with respect to $\Phi$ (only the dipole energy depends on $\Phi$) gives the equation
\begin{eqnarray}
\left[(1+\cos \Phi )u +\cos \Phi -{1\over 2}\right](1+u)\sin\Phi=0.
     \label{dip-phi}     \end{eqnarray}   
Solution of this equation yields
\begin{eqnarray}
\cos \Phi={1/2-u\over 1+u},~~v(u,\Phi) =0  
  \label{v0}  \end{eqnarray}
for $\beta<104^\circ~(u>-1/4)$  and
 \begin{eqnarray}
\cos \Phi=1,~~v(u,\Phi)=v_0(u) ={8\over 15 }\left({1\over 4}+u\right)^2  
    \label{dip}      \end{eqnarray}   
for $\beta>104^\circ~(u<-1/4)$.  
 
The spin-precession vortex state is non-uniform, and the gradient energy becomes essential. For an axially symmetric vortex with $2\pi$ circulation of the precession phase $\alpha$ ($\nabla \alpha=1/r$) the Euler-Lagrange equations are
\begin{eqnarray}
{1\over \xi_F^2}-{4- u \over 3r^2}
\nonumber \\
-{u\over (1-u^2)^2}\left(d u\over dr\right)^2
-{1 \over 1-u^2}\left({1\over r}{d u\over dr}+{d^2 u\over dr^2}\right)
\nonumber \\
+{4 \over 15\xi_d^2}\left[(1+\cos \Phi )u +\cos \Phi -{1\over 2}\right](1+\cos \Phi )=0,
  \label{u}        \end{eqnarray}   
\begin{eqnarray}
\left({1\over r}{d \Phi \over dr}+{d^2 \Phi\over dr^2}\right)
\nonumber \\
+{ 1\over 5\xi_d^2}\left[(1+\cos \Phi )u +\cos \Phi -{1\over 2}\right](1+u)\sin\Phi =0.
    \label{phi}      \end{eqnarray} 

There are two types of vortices corresponding to two types of symmetry. The solution with $\Phi=0$  is parity-symmetric, while in the structure with $\Phi \neq 0$ symmetry with respect to parity transformation $\Phi\to -\Phi$ is broken.  At the periphery of the vortex core, where $\Phi \ll 1$ and $u+1/4 \ll 1$, the solution of Eqs. (\ref{u}) and (\ref{phi}) is 
\begin{eqnarray}
u+{1\over 4} \approx  -{15\xi_d^2\over 16 \xi_F^2}+{85\xi_d^2\over 64  r^2} +{3\Phi^2 \over 16},
\nonumber \\
\Phi = C \sqrt{\xi_F \over r} e^{-3r/4\sqrt{2}\xi_F}.
\label{asym}  
        \end{eqnarray} 
where the constant $C$ is zero for the symmetric vortex and is of the order of unity for the non-symmetric vortex. The ratio between Fomin's and the dipole length is determined by the value of $u$ at infinity: $\xi_d=4 \xi_F \sqrt{|u_\infty+1/4|/15}$.  Equation (\ref{asym}) demonstrates that in the limit $\xi_d \ll \xi_F$ ($|u+1/4| \ll 1$) the first two terms in the expression for $u$ can be neglected everywhere except for very large distances $r$ of the order or larger than $\xi_F \ln (\xi_F/\xi_d)$. It is the approximation of Fomin, who used  the relation Eq.~(\ref{v0}) between $\Phi$ and $u=\cos \beta$ obtained for the uniform state for $u>-1/4$. This allows to reduce two coupled equations (\ref{u}) and (\ref{phi}) to a single one after exclusion from them the terms $\propto 1/\xi_d^2$. The resulting equation does not contain the dipole length explicitly, but the length $\xi_F$, which determines the core size,  certainly depends on it.

\begin{figure}[!b]
\begin{center}
   \leavevmode
  \includegraphics[width=0.9\linewidth]{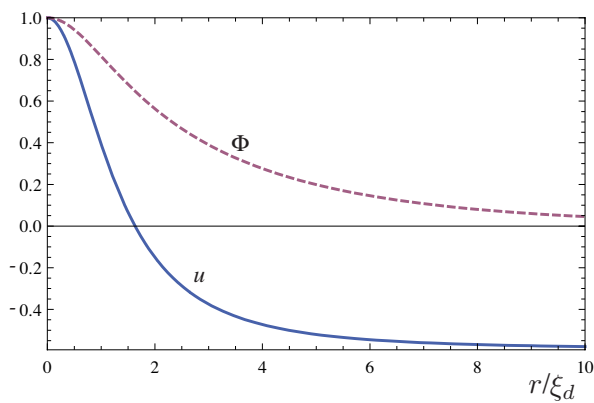}
 \caption{FIG. 1: Vortex structure at the transition from parity-symmetric to parity-nonsymmetric vortices. The plot of $\Phi$ (dashed line)  is normalized to the value $\Phi(0)=1$. }
 \label{fig1}
 \end{center}
\end{figure}

In order to find whether the symmetric $\Phi=0$ vortex can be realized we studied stability of the solution of Eq. (\ref{u}) for $\Phi=0$. This requires an analysis  of a linear equation for $\Phi$,
\begin{eqnarray}
-\left({1\over r}{d \Phi \over dr}+{d^2 \Phi\over dr^2}\right)
-{ 2(u +1/4)(1+u)\over 5\xi_d^2}\Phi = \epsilon \Phi,
 \label{phi-l}
          \end{eqnarray} 
which is analogous to the 2D Schr\"odinger equation for a particle in a potential well. Here $u$ is determined from Eq. (\ref{u}) at $\Phi=0$. The $\Phi=0$ vortex becomes unstable if Eq. (\ref{phi-l}) has a solution  with negative energy $\epsilon<0$. It is known from quantum mechanics that the 2D potential well has always a bound (localized) state\cite{LL}. But since for $u<-1/4$ (the necessary condition for stable spin-precession current) the ``potential energy'' $ 2(u +1/4)(1+u)/ 5\xi_d^2$ in Eq. (\ref{phi-l}) 
is positive at infinity,  the existence of a localized state leads to the condition $ \epsilon+ 2(u_\infty +1/4)(1+u_\infty)/ 5\xi_d^2<0$, which does not rules out that $\epsilon >0$, i.e., the solution $\Phi=0$ can be stable (except for the limits $u_\infty \to -1/4$ and $u_\infty \to -1$). The phase transition between two types of vortices is determined by the condition $ \epsilon=- 2(u_\infty +1/4)(1+u_\infty)/ 5\xi_d^2$. Studying Eq. (\ref{phi-l})  numerically we have found that $\epsilon$ reaches its critical value at $u_\infty = -0.594$, which corresponds to the precession angle $\beta=\arccos(u_\infty)=126.5^\circ$. Using the solution for $\Phi$ in the critical point for estimation of the energy contribution of the terms of order $\Phi^4$ we have found that this contribution is positive \cite{u}. Therefore the phase transition between the symmetric and the nonsymmetric vortices is of the second order. The numerical solutions of Eq. (\ref{u}) (with $\Phi=0$) and  Eq. (\ref{phi-l}) are shown in Fig.~1. 

We analyzed the spin-precession vortex in the absence of spin current at infinity. If an uniform spin current is present,  motion of vortices across current streamlines leads to precession-phase slips suppressing the current. If the precession-phase  gradient connected with the current exceeds the inverse core radius, there is no barrier impeding this process. The analysis of the Landau criterion showed\cite{son-87}  that the spin-precession current is stable up to the critical  gradient on the order of the inverse dipole length $1/\xi_d$.  In the limit $u\to -1/4$ the latter can essentially exceed the inverse core radius  $\sqrt{|u+1/4|}/\xi_d$.  But how can currents with gradients between $1/\xi_d$ and 
$\sqrt{|u+1/4|}/\xi_d$ be stabilized? The answer to this question is that the barrier blocking phase slips is present at the very early stage of nucleation of vortex rings in 3D systems (or vortex pairs in 2D systems). Vortex nucleation starts from a slight localized depression of the superfluid density (determined by $A(u)$ in our case). In order to analyze this "protonucleus" of the vortex core in the spin-current state one should consider a new Gibbs thermodynamic potential $\tilde G=G -\chi c_\perp^2 j \nabla \alpha/\gamma^2$:
\begin{eqnarray}
\tilde G={\chi c_\perp^2\over \gamma^2}\left[A(u){\nabla\alpha^2\over 2}+{\nabla u^2\over 2(1-u^2)} +{u\over \xi_F^2}
\right. \nonumber \\ \left.
+{v(u)\over \xi_d^2} -j\nabla \alpha \right],
        \end{eqnarray} 
where $j$ is a Lagrange multiplier.The nucleus, which is related with a peak of a barrier, corresponds to a saddle point in the functional space. So its structure should be found from  solution of  the Euler-Lagrange equations for the introduced Gibbs potential. The first step is to vary the Gibbs potential with respect to $\alpha$. Let us restrict ourselves with a 1D problem, when the distribution in the nucleus depends only on one coordinate $x$. Then the distribution of $\nabla \alpha$ is given by
$\nabla \alpha= j/A(u)$, where the reduced spin-precession current $j= A(u_\infty)\nabla  \alpha_0$ is determined by the gradient $\nabla  \alpha_0$ far from the nucleus center.  Expanding with respect to small deviation $g=u-u_\infty$ from the equilibrium value of $u$ at infinity one obtains 
\begin{eqnarray}
{\gamma^2\over \chi c_\perp^2}\tilde G=-{j^2\over 2A(u)}+{(\nabla u)^2\over 2(1-u^2)} +{u\over \xi_F^2}+{v(u)\over \xi_d^2}
\nonumber \\
\approx {(\nabla g)^2\over 2(1-u_\infty^2)}+g\left\{{d\over du}\left[-{j^2\over 2A(u_\infty)}+{v(u_\infty)\over \xi_d^2}\right]+{1\over \xi_F^2} \right\}
\nonumber \\ 
+{g^2\over 2}{d^2\over du^2}\left[-{j^2\over 2A(u_\infty)}+{v(u_\infty)\over \xi_d^2}\right]
+{g^3\over 6}{j^2\over 2}{d^3A(u_\infty)^{-1}\over du^3},
        \end{eqnarray} 
where we took into account that $d^3v(u)/du^3=0$. The linear in $g$ terms must vanish at the stationary current state. The term quadratic in $g$ determines the stability of the current state: it vanishes at the Landau critical current
\begin{eqnarray}
j_c= {1\over \xi_d^2}{d^2v(u_\infty)\over du^2}\left\{{d^2[A(u_\infty)^{-1}]\over du^2}\right\}^{-1},
        \end{eqnarray} 
which was derived in Ref. \cite{son-87}.  Considering the case of the current close to the critical value and using the Taylor expansion of $A^{-1}(u)$ around $u=-1/4$ one obtains
\begin{eqnarray}
G  = {16\chi c_\perp^2\over 15\gamma^2}\left[{(\nabla g)^2\over 2}  + a { g^2\over 2}-b{ g^3\over 6}\right],
        \end{eqnarray} 
where
\begin{eqnarray}
a=0.239(j_c^2-j^2),~~b=0.577j_c^2.
        \end{eqnarray} 
The Euler-Lagrange equation for this Gibbs potential, $-\Delta g=ag-bg^2/2=0$,  determines the distribution of $g$: 
\begin{eqnarray}
g=g_0 \left(1- \tanh^2{x\over r_p}\right), 
        \end{eqnarray} 
where $g_0= 3a/b=1.24 (j_c^2-j^2)/j_c^2$ is the value of $g$ in the nucleus center and $r_p=2/\sqrt{a}=4.1 \xi_d/\sqrt{j_c^2-j^2}$ is the nucleus size. The energy of the nucleus,
\begin{eqnarray}
\epsilon= {16\chi c_\perp^2S\over 15\gamma^2} \int_0^{3a/b}\sqrt{ag^2- {b g^3\over 3}} dg={64\chi c_\perp^2\over 25\gamma^2}{a^{5/2} \over b^2}S
\nonumber \\
=0.214S{\chi c_\perp^2\over \gamma^2}{(j_c^2-j^2)^{5/2} \over j_c^4},
  \label{energ}      \end{eqnarray} 
 determines the barrier for the process of the vortex core nucleation. Here $S$ is the cross-section area of the channel. Since in the limit $j\to j_c$ the nucleus size $r_p$ is divergent  our 1D description is always valid  close enough to the critical point, where  $r_p \gg \sqrt{S}$. When $r_p$ becomes smaller than the transverse size of the channel, one should consider the 3D or 2D (in the case of a thin layer) nucleus.  The first stage of this problem is to find the distribution of $\nabla \alpha$ from the continuity equation  $\vec \nabla [g \vec \nabla \alpha]=0$. Its solution demonstrates that outside the nucleus the distribution of $\nabla \alpha$ is the same as around the vortex ring (3D case) or the vortex dipole (2D case). In particular, in the 2D case
\begin{eqnarray}
\vec \nabla \alpha =\vec \nabla \alpha_0 -\int_0^\infty g(r_1)r_1^2\,dr_1 \left[ {\vec \nabla \alpha_0\over r^2} -{2\vec r (\vec r\cdot \vec \nabla \alpha_0)\over r^4}\right]. 
        \end{eqnarray} 
In contrast to the 1D case, the relation between $\nabla \alpha$ and $g$ is not local, so the following variation of the Gibbs potential with respect to $g$ leads to an integro-differential  equation. However on the basis of our solution of the 1D problem and using the scaling arguments one may conclude that the nucleus size can be roughly estimated from the expression   Eq. (\ref{energ}) for the 1D case with replacing $S$ by $r_p^2$ or by $r_pd$ for the 3D and the 2D case respectively ($d$ is the  thickness of the 2D layer). 

In our analysis we assumed that the value of the precession angle $\beta_\infty =\arccos u_\infty $ far from centers of a vortex or a nucleus was fixed and exceeded 104$^\circ$ ($u_\infty <-1/4$). Otherwise ($u_\infty >-1/4$)  the dipole energy vanishes,  and without dipole energy no stable current is possible. Meanwhile,  Fomin\cite{Fom-87} suggested that the spin-current can be stable  even if  $u>-1/4$ and the Landau criterion is violated. He argued that emission of spin waves, which comes into play after exceeding the Landau critical gradient, is not essential in the experimental conditions (see also a similar conclusion after Eq.  (2.39) in the  review by  Bunkov\cite{Bun}). This argument is conceptually inconsistent. If  the experimentalists observed ``dissipationless'' spin transport simply  because dissipation was weak,  it  would be ballistic  rather than superfluid transport. The essence of the phenomenon of superfluidity is not the absence of sources of dissipation, but ineffectiveness of this sources following from energetic and topological considerations. The Landau criterion is an absolutely necessary condition for superfluidity.  Fortunately for the superfluidity scenario in $^3$He-B, Fomin's estimation of the role of dissipation by spin-wave emission triggered by violation of the Landau criterion is not conclusive. He found that this dissipation  is weak compared to dissipation by spin diffusion.  But this is an argument in favor of importance rather than unimportance of the Landau criterion. Indeed, spin-diffusion, whatever high the diffusion coefficient could be, is ineffective in the subcritical regime, in which the gradient of  the ``chemical potential'' $\omega_P$ is absent. On the other hand, in the supercritical regime  the ``chemical potential'' is not constant anymore and this triggers the strong spin-diffusion mechanism of dissipation.

It is worthwhile to remind that the Landau criterion is a necessary but not sufficient condition for transport without dissipation. At  the Landau critical gradient (current) the current state ceases to be metastable because the barrier leading to  metastability vanishes. But it is well known that in reality current dissipation via phase slips is possible even in the presence of barriers (due to thermal fluctuations or quantum tunneling). Therefore, the present work addresses only ``ideal'' critical currents (the upper bound for them) leaving ``practical'' critical currents beyond the scope of the analysis.  

In summary, the Letter analyzed stability of spin precession currents in superfluid $^3$He-B  when the precession angle $\beta$ is very close to  $\beta_c=104^\circ$ and the spin-precession vortex has a  core of the radius $\sim \xi_d /\sqrt {\beta-\beta_c}$ much larger than the dipole length $\xi_d$. Though a barrier for motion of  these large-core vortices across the current streamlines disappears at rather small precession-phase gradients $\sim \sqrt {\beta-\beta_c}/\xi_d $, spin-precession currents remain stable up to much large gradients $\sim 1/\xi_d$, which were estimated from the Landau criterion\cite{son-87}. Stability of currents in this case is provided by barriers at very early stage of vortex-core nucleation.
It was also demonstrated that at the precession angle 126.5$^\circ$ there is a second-order phase transition between a parity-symmetric and parity-nonsymmetric spin-precession vortex cores. 

The author thanks I.\,A. Fomin and G.\,E. Volovik for critical comments, which influenced conclusions of the present Letter.


\begin{thebibliography}{99}

\bibitem{Usp} E.\,B. Sonin,  Superflows and superfluidity, Usp. Fiz. Nauk {\bf137}, 267 (1982) [Sov. Phys.--Usp. {\bf 25}, 409 (1982)].

\bibitem{Fom-91} I.\,A.~Fomin, Spin currents in superfluid $^3$He,  Physica B {\bf 169}, 153 (1991).


\bibitem{Bun} Yu.\,M.~Bunkov, Spin supercurrent and novel properties of NMR in $^3$He,
in  {\sl Progress of Low Temperature Physics}, vol. 14, edited by W.  P. Halperin, Elsevier, Amsterdam,  (1995),  p. 68.


\bibitem{n} G.\,E.~Volovik, Twenty years of magnon Bose condensation and spin current superfluidity in $^3$He-B,  cond-mat/0701180.

\bibitem{n1} Yu.\,M.~Bunkov and G.\,E.~Volovik, Magnon condensation into a $Q$ ball in $^3$He-B,  Phys. Rev. Lett. {\bf 98}, 265302 (2007).


\bibitem{n2} Yu.\,M.~Bunkov, Spin supercurrent, J. Magn. Magn. Mater. {\bf 310}, 1476 (2007).


\bibitem{n3} Yu.\,M.~Bunkov and G.\,E.~Volovik, Bose-Einstein condensation of magnons in superfluid $^3$He, J. Low Temp. Phys. {\bf 150}, 609 (2008).

\bibitem{n4} Yu.\,M.~Bunkov and G.\,E.~Volovik, Spin vortex in magnon BEC of superfluid  $^3$He-B,  Physica C {\bf 468}, 135 (2008).

\bibitem{A} T.~Sato, T.~Kunimatsu, K.~Izumina, A.~Matsubara, M.~Kubota, T.~Mizusaki, and Yu.\, M.~Bunkov, Observation of coherent precession of magnetization in superfluid  $^3$He A-phase,  arXiv: 0804.2994.

\bibitem{fn} As well as in Refs. \onlinecite{son-87,son-88} we prefer to tell about {\em spin-precession} rather than {\em spin} vortex since strictly speaking in $^3$He-B one deals not with currents of spin but with currents of precession moment (see further in the Letter).


\bibitem{son-78} E.\,B. Sonin, Analogs of superfluid currents for spins and electron-hole pairs, Zh. Eksp. Teor. Fiz. {\bf 74},
  2097 (1978)  [Sov. Phys.--JETP  {\bf 47},  1091 (1978)].
  
\bibitem{son-87} E.\,B. Sonin, Superfluid transport of precession in $^3$He-B,  Pis'ma Zh. Eksp. Teor. Fiz. {\bf 45},
 586 (1987)  [JETP Lett. {\bf 45},  747 (1987)].
 
 \bibitem{son-88} E.\,B. Sonin, Magnetic superfluidity in $^3$He, Zh. Eksp. Teor. Fiz.  {\bf 94},
100 (1988) [Sov. Phys.-JETP {\bf 67}, 1791 (1988)].
  
   \bibitem{Fom-88} I.\,A.~Fomin, Steady-state spin current in $^3$He-B, Zh. Eksp. Teor. Fiz. {\bf 94},
  112 (1988)  [Sov. Phys.--JETP  {\bf 67},  1148 (1988)].

\bibitem{misir} T.\,Sh. Misirpashaev and G.\,E. Volovik, Topology of coherent precession in superfluid $^3$He-B,  Zh. Eksp. Teor. Fiz.  {\bf 102}, 1197 (1992) [Sov. Phys.-JETP {\bf 75}, 650 (1992)].

\bibitem{PhTr} O.\,T. Ikkala, G.\,E. Volovik, P.\,J. Hakonen, Yu.\,M. Bunkov,
S.\,T. Islander, and G.\,A. Kharadze, NMR on rotating superfluid $^3$He-B, Pis'ma Zh. Eksp. Teor. Fiz. {\bf 35}, 338 (1982) [JETP Lett. {\bf 35}, 416 (1982)].

\bibitem{Thun} E.\,V. Thuneberg, Identification of vortices in superfluid $^3$He-B, Phys. Rev. Lett. {\bf 56}, 359 (1986).

\bibitem{Vol} M. M. Salomaa and G. E. Volovik, Vortices with spontaneously broken axisymmetry in $^3$He-B, Phys. Rev. Lett. {\bf 56}, 363 (1986).

\bibitem{Core} Y.  Kondo, J.\,S. Korhonen, M. Krusius, V.\,V. Dmitriev, Yu.\,M.
Mukharsky, E.\,B.  Sonin, and G.\,E. Volovik, Direct observation of the nonaxisymmetric vortex in superfluid $^3$He-B,  Phys. Rev. Lett. {\bf 67}, 81 (1991).

\bibitem{LT} A.\,J.~Leggett and S.~Takagi,  Orientational dynamics of superfluid $^3$He: A ``two-fluid'' model. I. Spin dynamics with relaxation, Ann. Phys. (N.~Y.) {\bf 106}, 79 (1977).


 \bibitem{Fom-87} I.\,A.~Fomin, Critical superfluid spin current in $^3$He-B, Pis'ma Zh. Eksp. Teor. Fiz. {\bf 45}, 106 (1987)  [JETP Lett. {\bf 45},  135 (1987)].

\bibitem{LL} L.\,D. Landau and E.\,M. Lifshitz, {\sl Quantum mechanics} (Pergamon Press, Oxford, 1977), Sec. 45.

\bibitem{u} Estimating the terms $\propto \Phi^4$ one must take into account  corrections of the order $\Phi^2$ to $u$.  

\end{thebibliography}
\end{document}